\documentclass[aps,prl,twocolumn,reprint,floatfix,showpacs]{revtex4}
\usepackage{graphicx}
\usepackage[pdftex,colorlinks,citecolor=blue,bookmarks]{hyperref}
\usepackage{bm}
\usepackage{longtable}
\usepackage{gensymb}
\date{\today}
\usepackage{color}
\begin{document}

\title{Collective and Single-particle Motion in Beyond Mean Field Approaches}
\author{J. Luis Egido}
\author{Marta Borrajo}
\author{Tom\'as R. Rodr\'iguez}
\affiliation{Departamento de F\'isica Te\'orica, Universidad
  Aut\'onoma de Madrid, E-28049 Madrid, Spain}


\begin{abstract}
We present a novel nuclear energy density functional method to calculate  spectroscopic properties of atomic nuclei. Intrinsic nuclear quadrupole deformations  and rotational frequencies are considered simultaneously as the degrees of freedom within a symmetry conserving configuration mixing framework. The present method allows the study of nuclear states with collective and single-particle character. We calculate the fascinating structure of the semi-magic $^{44}$S nucleus as a first application of the method, obtaining an excellent quantitative agreement both with the available experimental data and with state-of-the-art shell model calculations.
\end{abstract}

\maketitle
Mean field (MF) based methods~\cite{BHR.03} and the interacting shell model (SM)~\cite{Review_SM} are the cornerstones for the understanding of nuclear structure phenomena.  The traditional MF approach
 restricted to describe global properties of atomic nuclei has evolved  with the modern beyond mean field (BMF) methods to a more ambitious one, namely the study of the nuclear spectroscopy.

\begin{figure}[t]
{\centering
{\includegraphics[angle=0,width=.9\columnwidth]{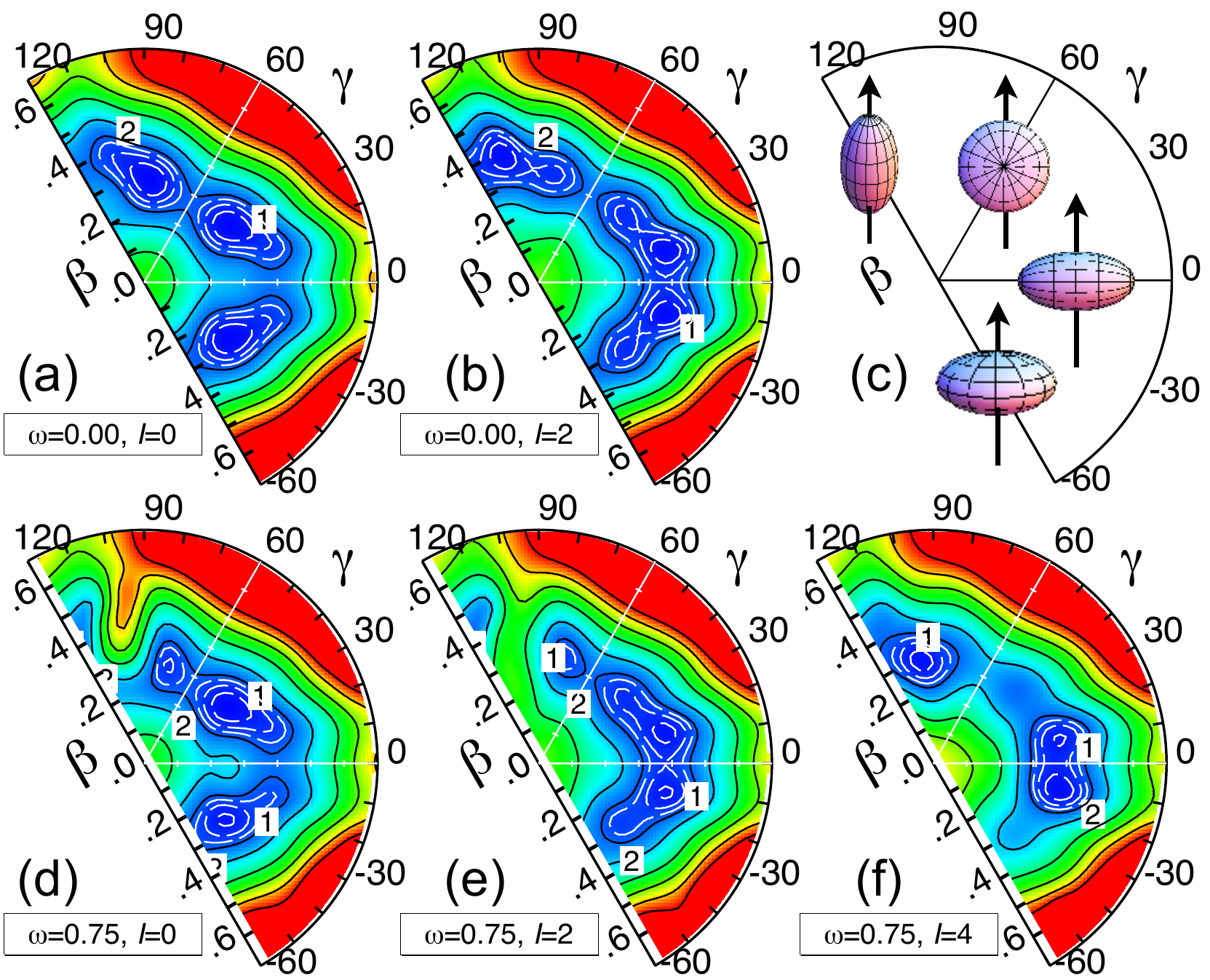}}
 \par}
\caption{(Color online) Potential energy surfaces in the $(\beta,\gamma)$ plane for two angular frequencies and three angular momenta for the nucleus $^{44}$S. The energy origin has been set at the energy minimum. The white dashed contours correspond to $0.25, 0.50$ and $0.75$ MeV, the 
unlabeled black contours start at 4 in steps of 2 MeV until 10 MeV. The units are: $\hbar \omega$ in MeV,  $I$ in $\hbar$ and $\gamma$ in degrees. }
\label{E_ome}
\end{figure}

 In the basic mean field approach, the Hartree-Fock-Bogoliubov (HFB) approach~\cite{RS.80} one can already cope with collective phenomena like the rotations or the superfluidity by the spontaneous symmetry breaking mechanism.  BMF approaches have mainly been developed either in small configuration spaces and using shell-model-like interactions \cite{SG.87,Sch.04} or in large configuration spaces and employing density-dependent interactions \cite{BHR.03} as discussed in this Letter.
Nevertheless, state-of-the-art BMF methods based on energy density functionals -Skyrme, Gogny and covariant density functionals- provide, in general, only a good qualitative agreement with the experimental  spectra. The  BMF methods have been developed in two directions: a) the recovery of the symmetries broken in the HFB approach, like particle number (PN) and angular momentum (AM) projection; and b) the incorporation of  fluctuations around the most probable MF values in the frame of the generator coordinate method (GCM). The combination of these two directions in a unified framework is the so-called symmetry conserving configuration mixing (SCCM) method.
The best current SCCM calculations~\cite{PRC_78_024309_2008,PRC_81_044311_2010,PRC_81_064323_2010} include the quadrupole (axial and triaxial) deformations as degrees of freedom and contain the AM projection within the projection after variation (PAV) approach~\cite{RS.80}. An awkward feature of the AM-PAV is a stretching of the whole spectrum \cite{PRL_99_062501_2007}. This is  related with the lack of an AM dependence in the variational equations to determine the HFB w.f. which favors  $I=0~\hbar$ states and disfavors the $I\ne0~\hbar$ ones (the larger $I$ the more).  In the past, AM dependence has been implemented  by the cranking technique which entails the time reversal symmetry breaking (TRSB) of the HFB w.f. and alignment. The suitability of this procedure has been shown in the cranked HF \cite{PRC_76_044304_2007} (HFB  \cite{HHR,ETY}) plus AM projection for Yrast states, and very recently in GCM calculations \cite{BRE.15} considering however only the collective sextant $0^{\circ}\leq \gamma \leq 60^{\circ}$ -see Fig.\ref{E_ome}(c)- and calculating only the $2^+_1$ and $4^+_1$ states of the Mg isotopes. 

In this Letter we push forward the state-of-the-art SCCM methods both by including the cranking frequency $\hbar \omega$ and by extending the range of triaxial quadrupole deformations to $-60^{\circ}\leq \gamma \leq 120^{\circ}$ to the triaxial GCM -see Fig.~\ref{E_ome}(c). These improvements not only largely solve the problems of the current BMF approaches but also include single-particle effects through the pair alignment by the cranking procedure.  
The success of the present approach can be understood from the shell model point of view if one considers that the wave function of a
deformed shape can be expanded as a linear combination of $n$-particle $n$-hole ($n$p-$n$h) excitations of the spherical mean field state \cite{ERR.04}, see also \cite{BAB.14} for odd-nuclei. Previous SCCM calculations were limited by construction to $n$p-$n$h excitations coupled to AM zero. The  consideration of the cranking frequency as 
coordinate opens the possibility of including $n$p-$n$h excitations coupled to AM different from zero making the variational space much richer.  Of course, the larger the number of generator coordinates considered the better is the approach.  Unfortunately  the CPU time needed for the calculations increase substantially with the number of coordinates.
To illustrate the new approach, we have chosen the exotic $N=28$ isotone $^{44}$S in which several unconventional properties have been  observed. The significant $2^+_1$ to $0^+_1$ transition probability~\cite{PLB_395_163_1997} suggests the erosion of the $N=28$ shell closure, the presence of a low-lying $0^+_2$  state~\cite{EPJA_25_111_2005,PRL_105_102501_2010} indicates  shape coexistence and the very low $4^+_1$ to $2^+_1$ transition probability suggests a $K=4$ isomeric state~\cite{PRC_83_061305_2011}. All these findings have motivated an unusual theoretical activity on this nucleus. There are mean-field calculations with Skyrme and relativistic interactions \cite{PLB_333_303_1994,PRC_60_014310_1999} and BMF studies with density functionals \cite{EPJA_9_35_2000,REM.02,RE.11,Li.11}  supporting the erosion of the $N=28$ shell closure and the manifestation of possible shape mixing and/or coexistence. 
Furthermore, large scale SM calculations have been performed~\cite{EPJA_25_111_2005,No.09,CG.14,Ka.13,Otska.15}  providing a good description of the data. Recently, the Tokyo group~\cite{Otska.15} proposed  a new type of high-$K$ isomerism to explain the long lifetime (of the order of 50 ps~\cite{PRC_83_061305_2011}) of the $4^{+}_{1}$ state.  This state and its associated band was not found in our earlier calculations~\cite{RE.11}. Thus, the calculations we present here are a good benchmark for our new theory. 

The nuclear w.f.s of the new approach have the form
\begin{eqnarray}
|\Phi^{I\sigma}_{M}  \rangle  &=&  \sum_{\lbrace\xi\rbrace}  f^{I\sigma}_{\lbrace\xi\rbrace}|IM;NZ;\lbrace\xi\rbrace \rangle
\label{GCM_ansatz}
\end{eqnarray}
where $\lbrace\xi\rbrace$ is the set of parameters $\lbrace\beta,\gamma;\omega;K\rbrace$ and $|IM;NZ;\lbrace\xi\rbrace\rangle=P^{Z}P^{N}P^{I}_{MK}|\phi(\beta,\gamma,\omega)\rangle$. These states are eigenstates of the symmetry operators. We  suppress the labels $N,Z$ hereafter to simplify the notation.
The operators $P^{Z}, P^{N}$ and $P^{I}_{MK}$ are projector operators associated with  the particle number and
the angular momentum, respectively, see \cite{PRC_81_064323_2010}, and $\sigma=1,2,...$ labels the states for a given value of the angular momentum $I$. The coefficients $f^{I\sigma}_{\lbrace\xi\rbrace}$ of the linear combination are found by a minimization of the energy in the Hilbert space spanned by the {\em linearly dependent} w.f.s $|IM;\lbrace\xi\rbrace\rangle$.  One obtains the  Hill-Wheeler equation
\begin{equation}
\sum_{\lbrace\xi\rbrace}\left(\mathcal{H}_{\lbrace\xi\rbrace,\lbrace\xi'\rbrace}^{I}-E^{I\sigma}\mathcal{N}_{\lbrace\xi\rbrace,\lbrace\xi'\rbrace}^{I}\right)f^{I\sigma}_{\lbrace\xi'\rbrace}=0.
\label{HWG_eq}
\end{equation}
Here we have introduced the  norm overlaps $\mathcal{N}^{I}_{\lbrace\xi\rbrace,\lbrace\xi'\rbrace} = \langle IM;\lbrace\xi\rbrace|IM;\lbrace\xi'\rbrace\rangle$ and the Hamiltonian overlap defined by a similar expression. 
Eq.~\ref{HWG_eq} is solved by standard techniques \cite{RS.80,PRC_78_024309_2008,PRC_81_064323_2010}: First, the  norm matrix is diagonalized, its eigenvalues $n^{I}_k$ and eigenvectors $u^{I}_k(\lbrace\xi\rbrace)$ provide the basis of the so-called natural states. 
The diagonalization of the Hamiltonian in this basis gives the eigenvalues $E^{I\sigma}$ of Eq.~\ref{HWG_eq} and the  eigenvectors $g_{k}^{I\sigma}$. In addition, the collective w.f.s $p^{I\sigma}(\beta,\gamma,\omega)=\sum_{k,K} g_{k}^{I\sigma} u^{I}_k(\lbrace\xi\rbrace)$ are orthogonal and $|p^{I\sigma} (\beta,\gamma,\omega)|^2$ can be interpreted as a probability amplitude. In the $(\beta,\gamma)$
plane the probability amplitude is defined by 
\begin{equation}
|{\cal P}^{I\sigma}(\beta,\gamma)|^2= \sum_\omega |p^{I\sigma} (\beta,\gamma,\omega)|^2.
\label{coll_wf}
\end{equation}
The HFB w.f.s $|\phi(\beta,\gamma,\omega)\rangle$ of Eq.~\ref{GCM_ansatz} are determined by minimizing the energy functional
\begin{equation}
 E[\phi]   =   \frac{ \langle \phi | HP^ZP^N|\phi \rangle } {\langle \phi | P^ZP^N|\phi \rangle }
 -  \langle \phi | \omega {\hat J}_x+ \lambda_{q_0} {\hat Q}_{20}+\lambda_{q_2 }{\hat Q}_{22}|\phi \rangle,  
   \label{cranking_1}
\end{equation}
where $\hat{Q}_{2\mu}$ and $\hat{J}_{x}$ are quadrupole moment and the $x$-component of the angular momentum operators respectively, $ \lambda_{q_0}$ and $\lambda_{q_2}$ the Lagrange multipliers determined by the constraints $\langle \phi |  {\hat Q}_{20}|  \phi \rangle = q_{20}$ and $\langle \phi |  {\hat Q}_{22} |  \phi \rangle =  q_{22}$, while $\omega$ is kept constant during the minimization process. $(\beta,\gamma)$ are defined by
$\beta= \sqrt{20\pi(q^{2}_{20} + 2 q^{2}_{22})}/3r^{2}_{0}A^{5/3}$ and $\gamma= \arctan(\sqrt{2}q_{22}/q_{20})$
with $r_{0}=1.2$ fm and  the mass number $A$.
That means, the HFB w.f.s are determined in the PN variation after projection (VAP) approach~\cite{NPA_696_467_2001}. Interestingly the incorporation of $\omega$ in the GCM Ansatz of Eq.~\ref{GCM_ansatz} is a generalization of the double projection method of Peierls and Thouless~\cite{PT.62,JLE.83} for the case of rotations. This method is known to provide the exact translational mass in the case of translations. We therefore expect that the moments of inertia of our bands will be close to the ones of the AM-VAP providing the sought after spectrum compression.
In the numerical applications the finite range density-dependent Gogny interaction with the D1S parametrization \cite{NPA_428_23_1984}  is used together with a configuration space of eight harmonic oscillator shells, large enough for realistic predictions for $^{44}$S. Concerning the generator coordinates we take three values of the angular frequency, namely, $\hbar \omega =0.0, 0.75$ and $1.25$ MeV, a discussion on this convergence will be given in Ref.~\cite{RBE.15}. For each $\hbar\omega$ value we take 70 points in the $(\beta,\gamma)$ plane, defined by $0\le \beta \le 0.7$ and $-60^{\circ}\le \gamma \le 120^{\circ}$ -see Fig.~\ref{E_ome}(c). We have to consider this larger $\gamma$ interval instead of the usual $0^{\circ}\le \gamma\le 60^{\circ}$ because, due to the term $-\omega {\hat J}_x$ in Eq.~\ref{cranking_1}, the HFB w.f. $|\phi\rangle$ is not time reversal invariant~\cite{BRE.15}.  These extensions increase drastically the computational burden, typically at least by two orders of magnitude.
We notice that rotations close to $\gamma= -60^{\circ}$ and $\gamma= 120^{\circ}$ are non-collective and can excite single particle degrees of freedom.  
\begin{figure}[t]
{\centering
 {\includegraphics[angle=-0,width=1.\columnwidth]{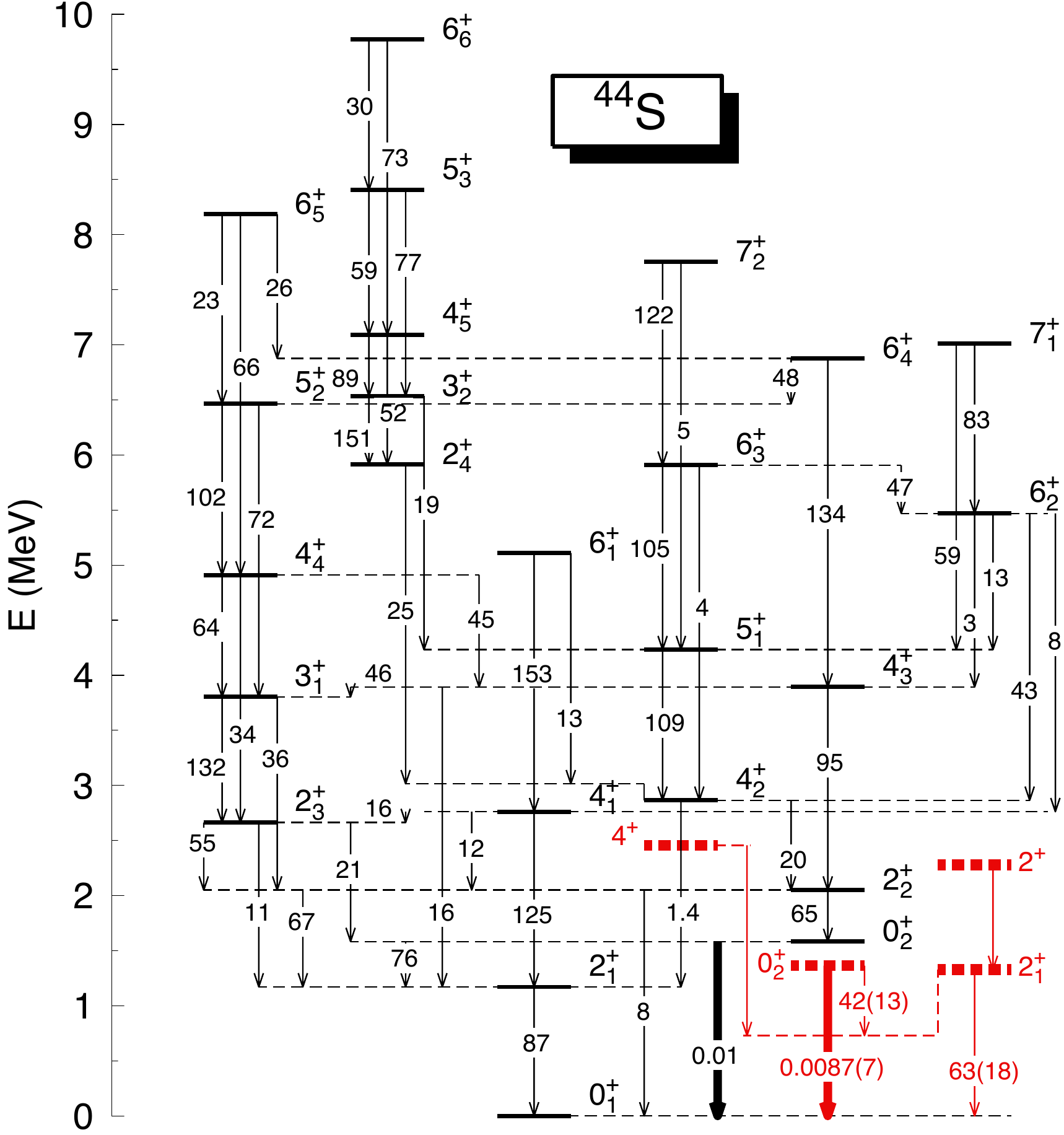}} \par}
\caption{Spectrum of $^{44}$S, showing the $B(E2)$ transition probabilities in e$^2$fm$^4$. The thick arrows represent the E0 transition with its corresponding value for $\rho^2$.  The experimental data~\cite{PRL_105_102501_2010,PRC_83_061305_2011} are  shown as thick dashed lines. Only experimental states with safe spin assignment are included.} \label{spectr_th}
\end{figure}

The GCM states -Eq.~\ref{GCM_ansatz}- recover the broken symmetries in the HFB approach and mixes different configurations $(\beta,\gamma,\omega)$, but one can also make a simplified Ansatz just fixing a given $(\beta,\gamma,\omega)$ value and mixing only in $K$ as to recover the symmetries, see for example Ref.~\cite{PRC_81_064323_2010}. In this case one can calculate the PN-AM projected energy in each point of the $(\beta,\gamma)$ plane and plot potential energy surfaces (PESs)  for different $\hbar\omega$ values. In Fig.~\ref{E_ome} we have represented these energies for $\hbar\omega =0.0$ and $0.75$ MeV and for $I=0, 2$ and $4~\hbar$. For $\hbar\omega=0.0$ MeV (Fig.~\ref{E_ome}(a)-(b)) we observe the mentioned symmetry, i.e., the three sextants are equivalent and can be obtained by reflexions around the axis $\gamma =0\degree$ and $\gamma =60\degree$. For  $I=0~\hbar$ we find a nucleus with $\beta\approx 0.30$ and very soft in $\gamma$, with a slight minimum at $\gamma\approx 30^{\circ}$. For $I=2~\hbar$ the lowest contours shifted towards the prolate and oblate shapes and somewhat larger $\beta$ values, and for $I=4~\hbar$, not shown here, the energy minimum close to the oblate shapes weakens about 1 MeV as compared with the prolate one. The effect of the angular frequency on the PESs can be seen in Fig.~\ref{E_ome}(d)-(f). We first observe that now the three sextants are not equivalent anymore. For $I=0 \; \hbar$ the PES looks similar to the case $\hbar \omega = 0$ MeV with the exception of the wedge around $\gamma=90 \degree$. For $I= 2 \; \hbar$ there are two minima at $\gamma \approx \pm 10\degree$ and at $\gamma \approx  \pm 45\degree$ and the wedge is still present. For $I=4\hbar$ and larger $I$-values the wedge disappears. The reason for this behavior is simple: For the $(\beta,\gamma)$ values inside the wedge, the HFB w.f. presents a neutron two-quasiparticle state with aligned AM,
$\langle \phi| \hat{J_x}|\phi \rangle\approx 4 \hbar$, making it costly to project to AM values smaller than 4$\hbar$. However, this is not the case for $I=4 \hbar$, Fig.~\ref{E_ome}(f), and we find three almost degenerated minima, two around
$\gamma\approx \pm 10\degree$ and $\beta= 0.35$ and a third one around $\gamma=90\degree$ and $\beta= 0.26$. 
The minima at $\gamma \approx 90^{\circ}$ and $\gamma \approx -45^{\circ}$  will play an important role in the interpretation of the collective w.f.s.
\begin{figure}[t]
{\centering
{\includegraphics[angle=0,width=.8\columnwidth]{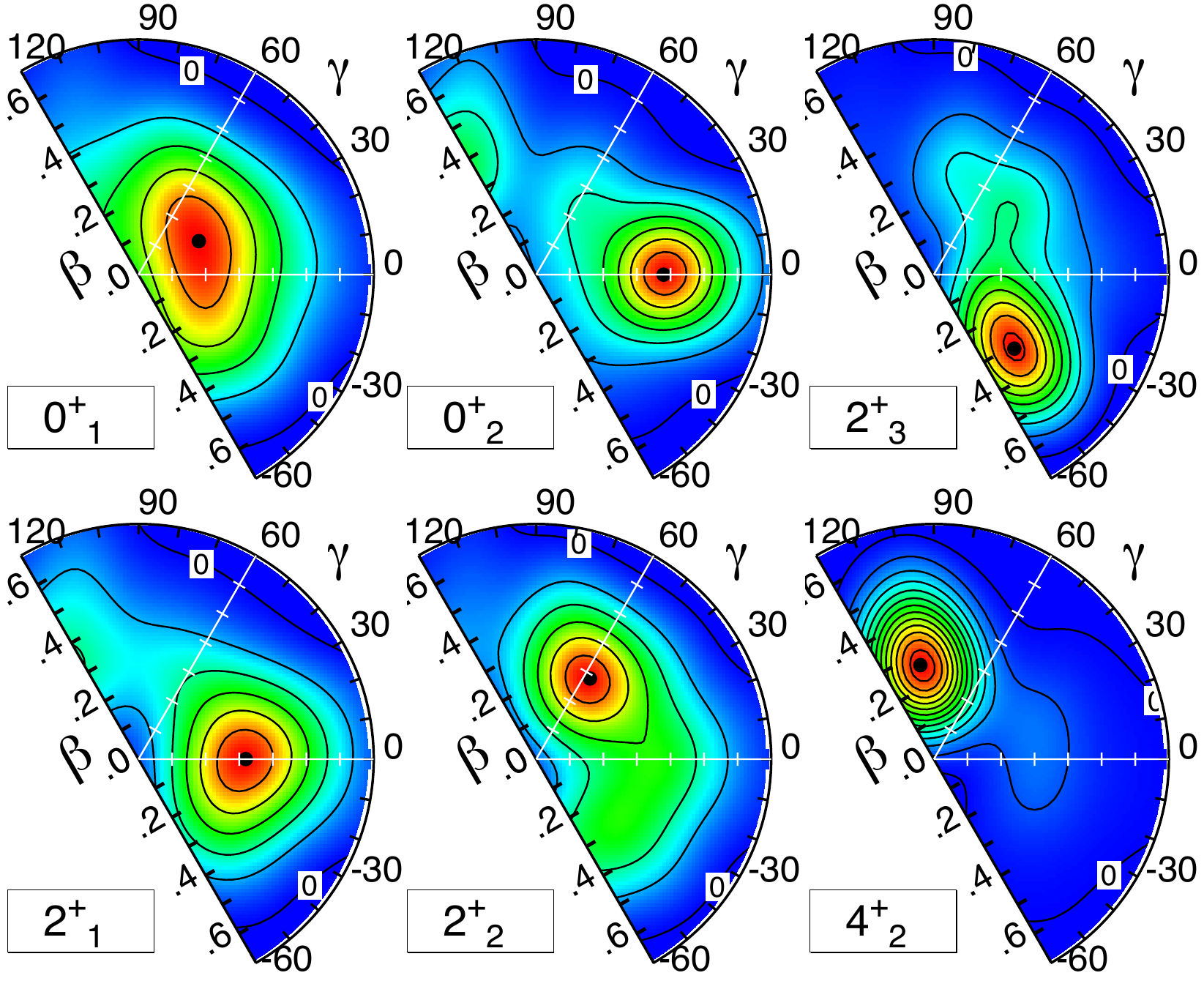}} \par}
 \caption{(Color online) Collective w.f.s in the $(\beta,\gamma)$ plane for the indicated states. The contour levels are
separated by 0.01. The contour labelled 0 sets the scale origin, the maximum is indicated by a black dot. }
\label{fig_coll_wf}
\end{figure}

The solution of Eq.~\ref{HWG_eq} provides the energy levels and the w.f.s. The transition probabilities~\cite{PRC_81_064323_2010,RBE.15} and the shapes of the w.f.s allow to order the energy levels into bands as shown in Fig.~\ref{spectr_th}. The lowest  levels provide the ground band, a band based on the $0^+_2$ level, two pseudo-$\gamma$-bands based on the $2^+_3$ and $2^+_4$ states, a band based on the $4^+_2$ level and a last one based on the $6^+_2$ state. For the physical interpretation of these bands we show in Fig.~\ref{fig_coll_wf} the collective w.f.s, see  Eq.~\ref{coll_wf}, of representative states.  
The minima of Fig.~\ref{E_ome} represent the relevant configurations and play a relevant role in the shape of the collective w.f.s.  
The high-$I$ members of a band with a w.f. looking similar to the band head are not plotted. The $0^+_1$ state presents a very extended  w.f. with contributions from many configurations and a maximum in the area $0^\circ \leq \gamma\leq 60^\circ$ and $0.15 \leq \beta \leq 0.3$. It resembles the PES of Fig.~\ref{E_ome}(a)-(b). The higher AM members of the band become prolate as can be seen in the w.f. of the $2^+_1$ state. The $0^+_2$ state, band head of the first excited band, is soft in the $\gamma$ direction and peaks at a prolate shape. The higher AM members of the band, however, are oblate, see for example the $2^{+}_{2}$ state in Fig.~\ref{fig_coll_wf}. The second excited band, based on the $2^+_3$ state presents a triaxial-oblate shape with the maximum at $\beta =0.32$ and $\gamma = -45^{\circ}$. The third, fourth and fifth  excited bands, with  the $4^+_2$, $2^+_4$ and $6^+_2$ states  as band heads, have maxima at  $\beta \approx 0.28-0.36$ and $\gamma \approx 90^{\circ}-100^{\circ}$, cf. the minimum at this point of Fig.~\ref{E_ome}(f).  Since the w.f.s of these three states look rather similar we only display the one of the $4^+_2$ state. The w.f. of the 
 $4^+_2$ state strongly peaks at the maximum indicating a less collective character. If we analyze the composition of the HFB w.f. at the maximum we find that it corresponds to an aligned state with contributions from the $\nu$f$_{7/2}$ and $\nu$p$_{3/2}$ orbitals. The band starting at this level has been assigned in Ref.~\cite{Otska.15} as a $K=4$ band. In the present calculations, with explicit breaking of the time reversal symmetry, the $K$ quantum number looses relevance.  However, in some cases, through the cranking mechanism, one has alignment along the $x-$axis which can be used instead to characterize bands. If we express the w.f. in the basis $|I K_X\rangle$,  with $K_X$ the projection of the angular momentum along the intrinsic x-axis, we obtain that the w.f. of this state is predominantly $K_X=4$, in agreement with the interpretation of Ref.~\cite{Otska.15}.
The band based on the $6^+_2$ level, is very similar to the one of the  $4^+_2$ state. In the basis $|I K_X\rangle$ the component with  $K_X=6$ amounts to $76\%$.  We would like to stress the special role played by the sextants 
$(0^{\circ},60^ {\circ})$ and $(60^{\circ},120^ {\circ})$ of the $(\beta,\gamma)$ plane. They provide new states, like the $4^+_2$, and contribute actively to the configuration mixing of other states.
The spherical configurations, not shown here, appear at several MeV of excitation energy, the lowest ones corresponding to the $0^+_3, 2^+_6, 4^+_8,$ and $6^+_8$ states, a clear indication of the erosion of the $N=28$ shell closure. The spectroscopic quadrupole moments of the band heads are~:
$Q_{spec}(2^+_1) = -14.4$ efm$^2$, $Q_{spec}(2^+_2) = 6.5$ efm$^2$,  $Q_{spec}(4^+_2) = 26.9$ efm$^2$, $Q_{spec}(2^+_3) = -13.8$ efm$^2$.
\begin{figure}[t]
{\centering
{\includegraphics[angle=0,width=\columnwidth]{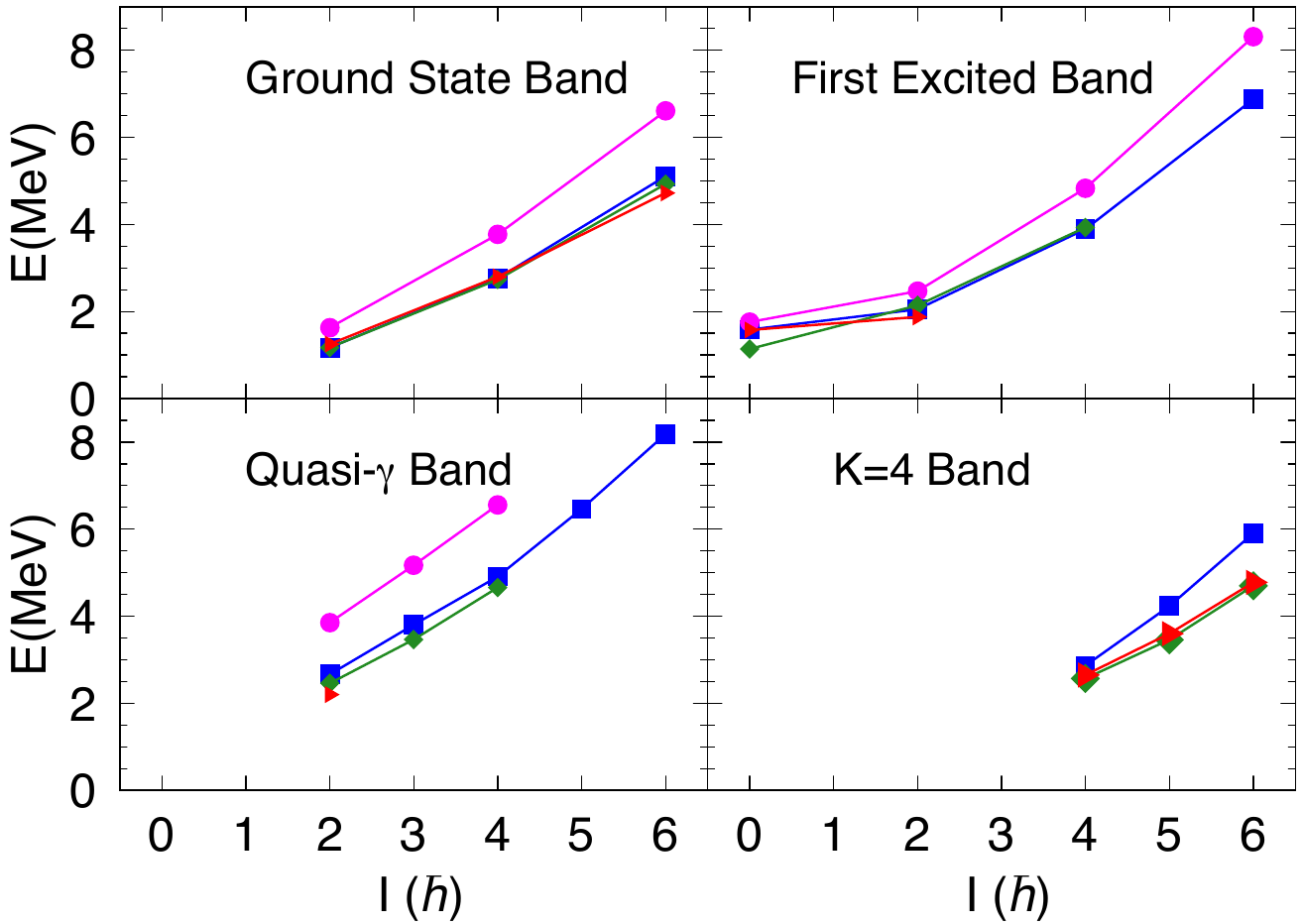}} \par}
 \caption{(Color online) Comparison of several theories: Triangles, red lines, Tokyo group \cite{Otska.15}; diamonds,  green lines, Madrid-Strasbourg collaboration \cite{PRC_85_024311_2012}; boxes,  blue lines, this work; circles,  magenta lines, our former work without angular frequency dependence \cite{RE.11}.}
\label{fig_theory_comp}
\end{figure}
For comparison the experimental data have also been plotted in Fig.~\ref{spectr_th} as thick lines. With respect to the energy values we obtain a good agreement. Concerning the transition probabilities very good agreement is found for the E0 from the $0^+_2$ to the $0^+_1$ state and the E2 from the  $2^+_1$ to the $0^+_1$ while the $B(E2; 0^+_2 \rightarrow 2^+_1)$ is slightly overestimated. In our calculations the $4^+_2$ state decays both to the $2^+_1$ (with a $B(E2)=1.4$ e$^2$fm$^4$) and $2^+_2$ states (with a $B(E2)= 20$ e$^2$fm$^4$). The latter decay branch has not been observed experimentally. Considering the theoretical values, we estimate a branching ratio of $74\%$  for the decay branch to the $2^+_1$ state and a lifetime of 84 ps to be compared with the experimental value of about 50 ps~\cite{PRC_83_061305_2011}. 
Another interesting finding is that the $6^+_2$ level, which is  similar in structure to  the  $4^+_2$ state, has a much shorter lifetime since it has several decay branches. Furthermore its small excitation energy above Yrast makes it experimentally accessible.

In Fig.~\ref{fig_theory_comp} we now compare the performance of the present method with state-of-the-art SM calculations of the Madrid-Strasbourg collaboration \cite{PRC_83_061305_2011} in the full sd(fp) valence space for protons (neutrons) with the SDPF-U interaction \cite{No.09,PRC_85_024311_2012} and with those of the Tokyo group \cite{Otska.15} in the $\pi$(sd)$^{(Z-8)}\nu$(pf)$^{(N-20)}$ and the SDPF-MU. The agreement between the two SM calculations and our present approach for the ground state and first excited bands is extraordinary. Also for the quasi-$\gamma$-band we find good agreement between our approach and the one of the Madrid-Strasbourg collaboration. Small deviations are observed for the $I=5 \hbar$ and $6\hbar$ states of the "$K=4$" band. The transition probabilities are also similar.  For example with the SDPF-U interaction one obtains~\cite{CG.14} $B(E2; 6^+_1 \rightarrow 4^+_1)=118$  e$^2$fm$^4$,  $B(E2; 4^+_1 \rightarrow 2^+_1)=111$ e$^2$fm$^4$,  
 $B(E2; 2^+_1 \rightarrow  0^+_1)=75$ e$^2$fm$^4$, to compare with our values of $153, 125$ and  $87$ e$^2$fm$^4$, respectively. We note that in our calculations no effective charges are used and that the D1S parametrization  was fitted long ago to provide reliable global properties along the nuclide chart, reinforcing the predictive power of our approach. 
In  Fig.~\ref{fig_theory_comp} we can also observe the improvement provided by the present approach as compared to our former results~\cite{RE.11}  obtained without considering the $\omega$ degree of freedom. These calculations gave the right tendency but an stretched spectrum which is corrected in the present framework (see also~\cite{BRE.15}).  Furthermore, we also observe that the aligned structures observed in the present calculations cause a decrease in the collectivity of the w.f.s and consequently a decrease of the transition probabilities which often were found too large in the past.  All these facts improve considerably the agreement of the present approach with the experiment. 
    
In conclusion, in this Letter we report on the consideration of cranked w.f.s together with triaxial deformations $(\beta,\gamma)$ in the Symmetry Conserving Configuration Mixing approach. The cranking procedure introduces an angular momentum dependence in the calculations providing a compression of the otherwise stretched spectrum. Furthermore, through the alignment mechanism, single particle degrees of freedom are introduced, opening a door to a physics unaccessible before in these approaches.  The aligned configurations provide a decrease of the collectivity of the w.f.s leading to smaller transition probabilities  in agreement with the experiment.  These three facts cure the deficiencies of former SCCM approaches providing a very powerful tool in nuclear structure calculations. In our example of the exotic nucleus $^{44}$S, with a very rich nuclear structure, we have shown that this approach provides high quality nuclear spectroscopy comparable with the state-of-the-art of SM calculations with tailored interactions. The advantages of our approach are the added value of the intrinsic system interpretation and that our interaction, the Gogny force, is well known for its predictive power and good performance for bulk properties all over the chart of nuclides. These calculations set a new standard in the state-of-the-art of BMF methods with density dependent interactions. 
A drawback of our approach in its present form is that the increase from one to tree sextants as well as the consideration of one more coordinate enlarge considerably the CPU time of the calculation.  Systematic studies or calculations with a very large number of major shells are not feasible in a small local cluster. We are currently studying different ways to speed up the calculations.

This	work	was	supported	by the Ministerio de Econom\'ia y Competitividad under contracts FPA2011-29854-C04-04, FPA2014-57196-C5-2-P, BES-2012-059405 and Programa Ram\'on y Cajal 2012 number 11420.

\end{document}